\documentclass[pdflatex,sn-mathphys]{sn-jnl}% Math and Physical Sciences Reference Style

\jyear{2021}%

\theoremstyle{thmstyleone}%
\theoremstyle{thmstyletwo}%

\theoremstyle{thmstylethree}%

\usepackage{comment}
\usepackage{lineno}

\graphicspath{}
\begin{document}

%\linenumbers

\title[Moments of multiplicity distributions]{
Moments of multiplicity distributions for KNO scaling  study  using   the ATLAS  results 
}
 

\author*[1,2]{\fnm{Yuri A.} \sur{Kulchitsky}}\email{Yuri.Koultchitski@cern.ch}

\author[1,2]{\fnm{Pavel} \sur{Tsiareshka}}
%\email{Pavel.Tsiareshka@cern.ch}


%\affil*[1]{ \orgname{Joint Institute for Nuclear Research}, \orgaddress{\street{ Joliot-Curie 6}, \city{Dubna}, \postcode{141980}, \country{Russia}}}
\affil[1]{ \orgname{Joint Institute for Nuclear Research}, \city{Dubna}, \country{Russia}}
\affil[2]{ \orgname{B.I.~Stepanov Institute of Physics, National Academy of Sciences}, \city{Minsk}, \country{Belarus}}

\abstract{
The normalised order-\(q\) moments of  primarily charged-particle multiplicity distributions
are studied  for KNO scaling investigation  in   \(pp\)  collisions    as deduced from 
the results of the ATLAS  at the LHC.
The  normalised moments  for  the LHC  and low-energy experiments  are compared  for 
the kinematic region with  an absolute pseudorapidity less than \(2.5\). 
The  normalised  moments show a  
%linearly  increase 
increases linearly 
with centre-of-mass energies,  and
therefore  the KNO scaling  is violated for  the  full-scaled   multiplicity region. 
The normalised moments for scaled multiplicity 
%more 
greater
than one average multiplicity
are constant for the highest  centre-of-mass energies,  and therefore  the KNO scaling is concluded to hold.
}


\maketitle
 
%\linenumbers
 
The investigation of charged-particle distributions in proton-proton (\(pp\)) collisions  probes the strong interaction  at the low-momentum transfer (non-perturbative region of  quantum chromodynamics).
The study of normalised order-\(q\) moments,  \( C_{\mathrm{q}} (\sqrt{s}) \),
of primary charged-particle multiplicity distributions  is sensitive to the KNO scaling. 
%
%__________________________________________________________________________________________
% YK
%The review of  the  KNO scaling and normalised order-\(q\) moments results
%for pp-interaction at the LHC  and lower energy experiments  was presented in Ref.\ \cite{Kulchitsky:2023tcq}. 
%__________________________________________________________________________________________

The KNO scaling hypothesis  means  that  at  energy  asymptotic,
the probability distributions   \( P (n, \sqrt{s}) \)  of producing \( n \)  particles in a certain collision
process should  demonstrate  a scaling relation  \cite{Polyakov:1970lyy,Koba:1972ng,Proceedings:1973mza}.
The  main assumption of  the KNO scaling is Feynman scaling    \cite{Feynman:1969ej},
where  it was  concluded that for asymptotically large  centre-of-mass (CM) energies with   
\(\sqrt{s} \rightarrow  \infty \)  the mean total number of any kind of particle 
logarithmically  rises with   the CM energy as   \(  \langle n \rangle  \propto \ln{\sqrt{s}}\).
For this assumption, 
the multiplicity distribution  \( P (n, \sqrt{s}) \)  was represented as
\begin{equation}
\label{eq_pn2}
P (n, \sqrt{s}) =  \frac{1}{\langle n (\sqrt{s}) \rangle} \Psi ( z )  + 
\mathcal{O} \left( \frac{1}{\langle n (\sqrt{s}) \rangle^{2}} \right) ,
\end{equation}
where   \( \langle n (\sqrt{s}) \rangle \)   is the average multiplicity of primary particles at  
the CM energy,   \( \Psi ( z ) \)  is  the particle distribution as a function of the scaled multiplicity
\( z =  { n (\sqrt{s}) }/{ \langle n (\sqrt{s}) \rangle } \). 
The first term in (\ref{eq_pn2}) results from the leading term in  \( \ln{\sqrt{s}}\) (KNO scaling hypothesis),   
and the second term contains all other terms \cite{Grosse-Oetringhaus:2009eis,Kittel:2005fu}. 
The multiplicity distributions become simple rescaled copies of the universal function   \( \Psi ( z ) \), 
depending only on the scaled multiplicity or an energy-independent function.
Asymptotically, 
when \(\sqrt{s} \rightarrow  \infty \),  the second term  in  (\ref{eq_pn2}) tends to zero, 
and therefore  the KNO scaling holds.

To precisely  find the KNO scaling, it is important to study the normalised order-\(q\) moments
of   primary charged-particle multiplicity distributions 
\begin{equation}
%\nonumber
\label{eq_Cq}
 C_{\mathrm{q}} (\sqrt{s}) = 
 \frac{ \langle n^q (\sqrt{s}) \rangle }{ \langle n (\sqrt{s}) \rangle^\mathrm{q} } ,
\end{equation}
where \(q\) is the order of the moment. 
%
%
%__________________________________________________________________________________________
% YK
The moments \( C_{\mathrm{q}} (\sqrt{s}) \) give an integral characteristic of multiplicity distributions \( P (n, \sqrt{s})  \), 
%when 
while
%as 
the
KNO scaling distributions, \( \Psi ( z )\),  give differential dependence from normalised multiplicity. 
The energy independence of  \( C_{\mathrm{q}} (\sqrt{s})  \)  of multiplicity distributions 
of various orders would imply  observation of  the KNO scaling.
Therefore, the investigation of \( C_{\mathrm{q}} (\sqrt{s})  \)  moments is
an 
absolutely 
essential addition to the study of  \( \Psi ( z ) \) functions.
%__________________________________________________________________________________________

In this paper,
the   normalised order-\(q\) moments  are studied using  the primary charged-particle 
multiplicity distributions \cite{STDM-2010-01,STDM-2010-06,STDM-2014-19,STDM-2015-02,STDM-2015-17}
by ATLAS \cite{PERF-2007-01}  at the LHC \cite{Evans:2008zzb}.
Measurements of the  primary charged-particle distributions in ATLAS at 
\(\sqrt{s}=0.9\),  \(2.36\), \(7\),  \(8\), 
and \(13\)~TeV  
were performed for  the  pseudorapidity region less than \(2.5\) 
and  for two samples of events: with 
the
primary charged-particle multiplicity
greater
%more 
than  or equal to \(2\) and \(1\)  and with  the charged-particle  transverse momentum 
\(p_{\mathrm{T}}\)  
greater
%more  
than \(100\) and \(500\)~MeV,  respectively. 
The study of the KNO scaling using 
the
\( \Psi ( z ) \) scaled multiplicity function, 
which is defined  in (\ref{eq_pn2})  
and calculated  on the ATLAS data
\cite{STDM-2010-01,STDM-2010-06,STDM-2014-19,STDM-2015-02,STDM-2015-17},
was published in  \cite{Kulchitsky:2022gkm} by these authors.
%__________________________________________________________________________________________
% YK
A comparison of \( C_{\mathrm{q}} (\sqrt{s}) \) with  the results of the LHC  and lower energy experiments 
for \( pp\) interactions is presented. 
%__________________________________________________________________________________________

%__________________________________________________________________________________________
% YK  
The KNO scaling and \( C_{\mathrm{q}} (\sqrt{s}) \)-moments were studied by the CMS 
at \( \sqrt{s} \) from \(0.9\)  to  \(7\)~TeV  in central pseudorapidity  \( \mid\eta\mid < 0.5 \) 
and more inclusive  \( \mid\eta\mid < 2.4 \) regions \cite{CMS:2010qvf}.
The KNO results were investigated by ALICE at \( \sqrt{s} \) from \(0.9\)  to \(13\)~TeV
in pseudorapidity regions 
% \( \mid\eta\mid < 0.5,\ 0.8,\ 1.0,\ 1.5 \) 
 \( \mid\eta\mid < 0.5 \),   \( \mid\eta\mid < 0.8 \),  \( \mid\eta\mid < 1.0 \),   
 and  \( \mid\eta\mid < 1.5\)
\cite{ALICE:2010cin,ALICE:2015olq,ALICE:2022xip,Fan:2022bbp}.
For ALICE and CMS  experiments the KNO scaling  is  violated for energies from  \(0.9\) to \(8\)~TeV
if taking into account more inclusive pseoudorapidity regions
because  the \(C_{\mathrm{q}}(\sqrt{s}) \)-moments are increase with energy. 
The KNO scaling holds  for the central pseudorapidity region with \(\mid\eta\mid < 0.5 \)
and for the energy region from \(\sqrt{s} = 0.9\) to \(7\)~TeV on the CMS data 
\cite{CMS:2010qvf}.
In this case, 
the \(C_{\mathrm{q}}(\sqrt{s}) \)-moments are independent of  energy.
The  results of ALICE for \(C_{\mathrm{q}}(\sqrt{s}) \)-moments in central pseudorapidity,  \(\mid\eta\mid < 0.5 \), 
are slightly increased with energy for \( q=4,\ 5 \). 
%__________________________________________________________________________________________

%_______________________________________________________________________
% Fig 1
%_______________________________________________________________________
\begin{figure*}[t!]
\centering
\begin{minipage}[h]{0.32\textwidth} 
\center{\includegraphics[width=1.0\linewidth]{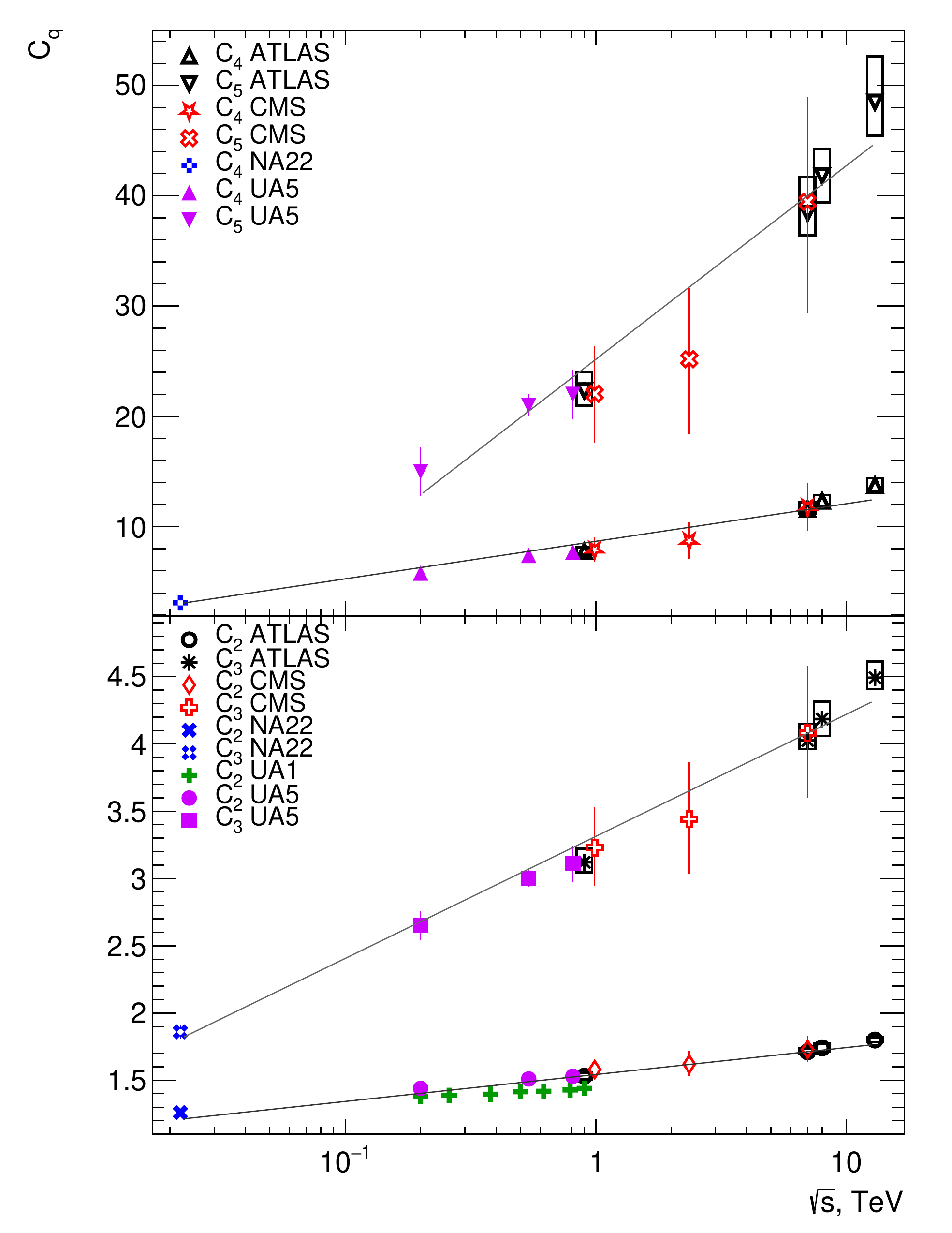}}  
(a)
\\
\end{minipage}
\hfill
%\hspace{2mm}
\begin{minipage}[h]{0.32\textwidth} 
\center{\includegraphics[width=1.0\linewidth]{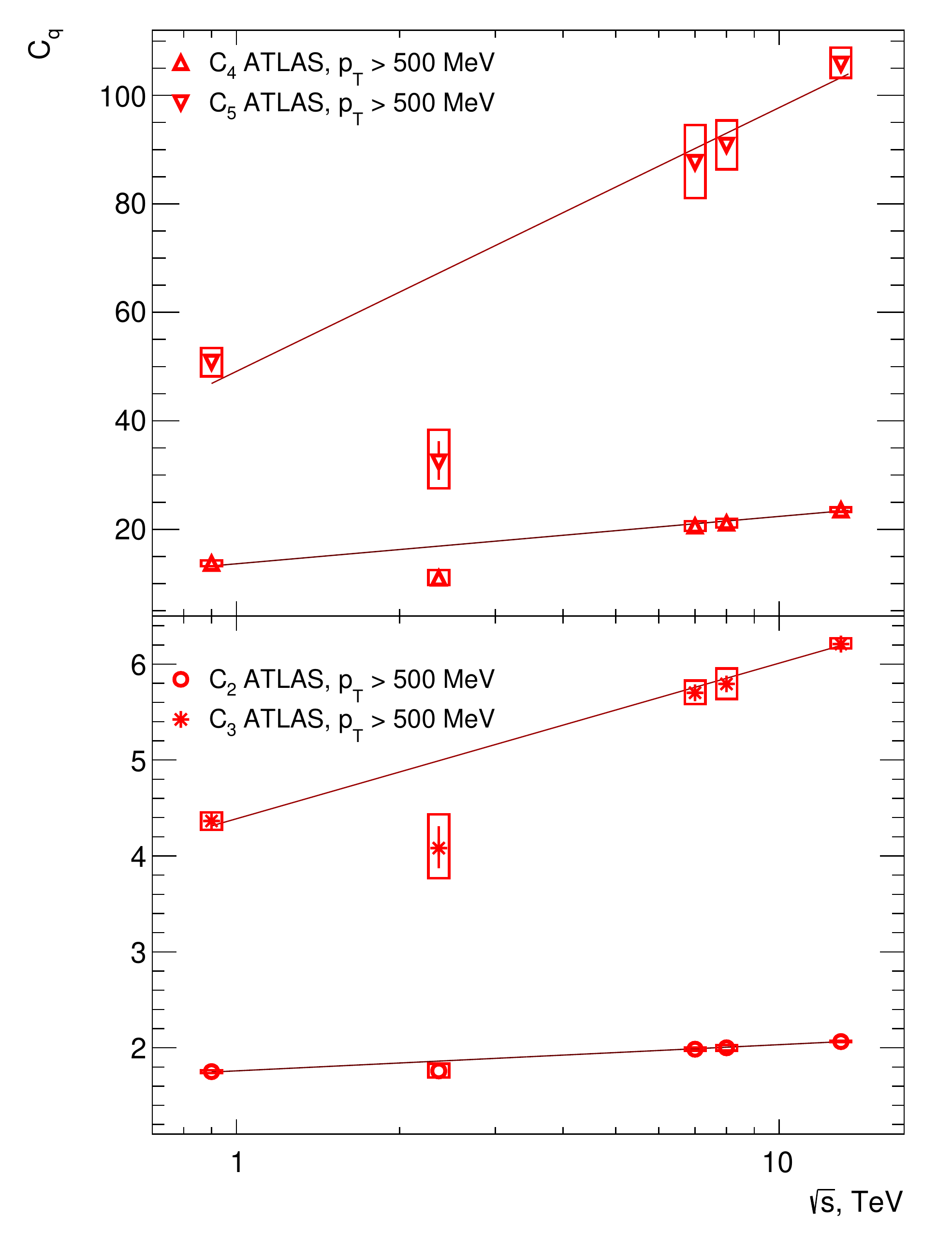}}  
(b)
\\
\end{minipage}
\hfill
\begin{minipage}[h]{0.32\textwidth} 
\center{\includegraphics[width=1.0\linewidth]{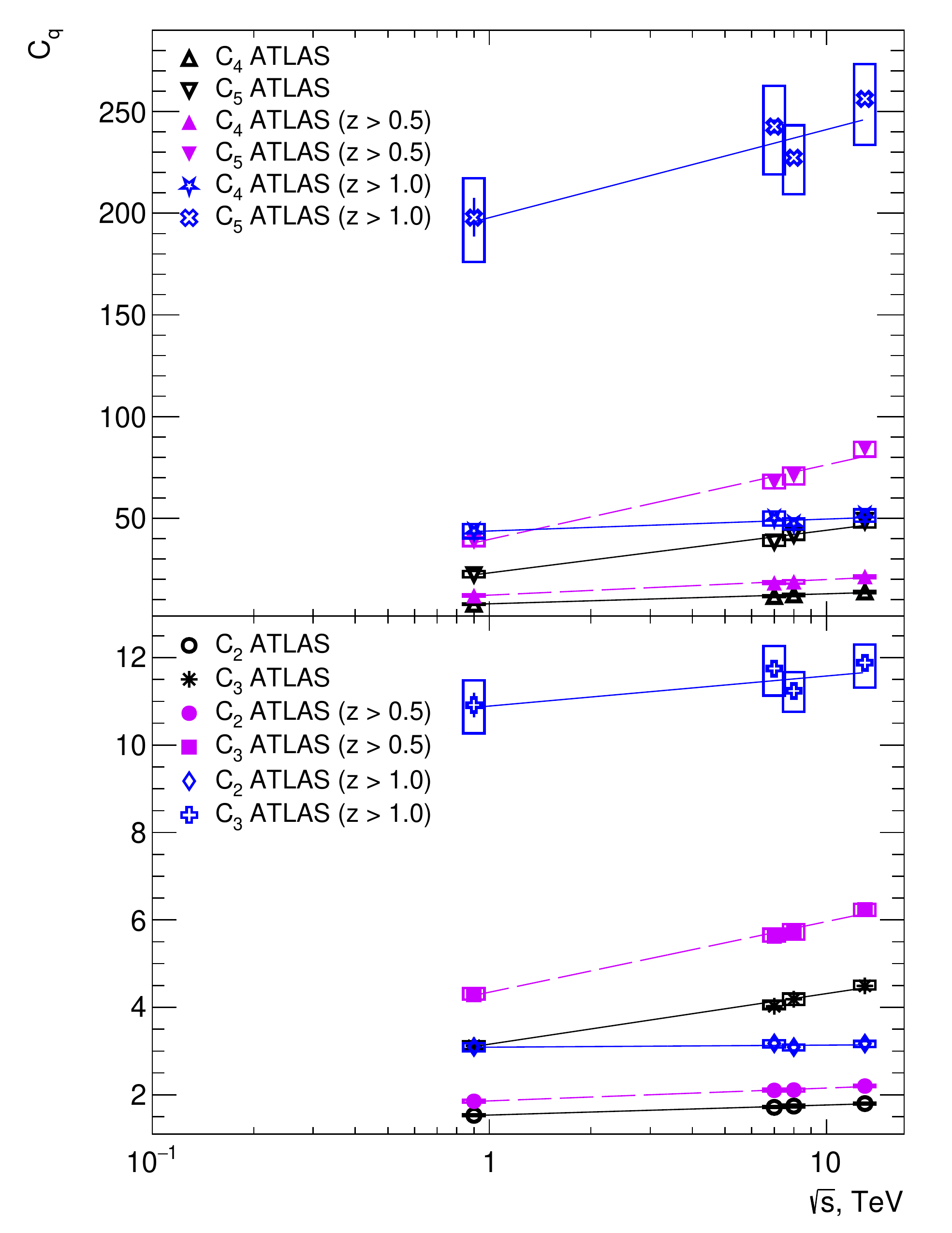}}  
(c)
\\
\end{minipage}
\caption{
The  normalised order-\(q\) moments,   \(C_{\mathrm{q}}(\sqrt{s})\),  in Eq.~(\ref{eq_Cq}) 
of the primary charged-particle multiplicity distributions   measured by 
ATLAS  for events  collected at \( \sqrt{s} = 0.9,\ 2.36,\ 7, 8\), 
and \(13\)~TeV for 
(a)
the  pseudorapidity region  \( \mid\eta\mid < 2.5\).  
The results of   CMS   \cite{CMS:2010qvf} and lower-energy experiments  NA22  
\cite{EHSNA22:1987syd},
UA1  
\cite{UA1:1989bou},  
and 
UA5 
\cite{UA5:1985hzd,UA5:1988gup}  are included. 
(b) 
The ATLAS results for
\( \mid\eta\mid < 2.5\), \(n_{\mathrm{ch}} \ge 1\),  \(p_{\mathrm{T}} >500\)~MeV.
(c) 
The ATLAS results for
\( \mid\eta\mid < 2.5\), \(n_{\mathrm{ch}} \ge 2\), \(p_{\mathrm{T}} >100\)~MeV
with additional scaled multiplicity thresholds:  \( z  > 0.5 \) and  \( z > 1.0 \).
The  
\(C_{\mathrm{2}}\),   
\(C_{\mathrm{3}}\)   
and  
\(C_{\mathrm{4}}\),
 \(C_{\mathrm{5}}\)  
results are shown in the  bottom and top panels,  respectively. 
The vertical bars  are the statistical uncertainties, and  the  squares  are the systematic uncertainties.
The coloured symbols  are the data. 
Fits of the  \( \log{\sqrt{s}}\)   dependence of the \( C_{\mathrm{q}} (\sqrt{s}) \) 
of the multiplicity distribution   (assuming linear dependence)   are shown.
In (a), 
for \(\sqrt{s} = 0.9\)~TeV,    
%data from experiments other than ATLAS were shifted drawn to  lower  \(\sqrt{s}\) for clarity.
data from non-ATLAS experiments have been shifted down \(\sqrt{s}\) for clarity.
The  lines show the results of the fits for  \(C_{\mathrm{q}}  (\sqrt{s}) \)  with statistical and systematic uncertainties 
added in quadrature.
}
\label{fig_Cq_ATLAS_1}
\end{figure*}
%_______________________________________________________________________

The results of this analysis
%This analysis results for
of the validity of KNO scaling  
%is 
are
shown quantitatively in   
Fig.~\ref{fig_Cq_ATLAS_1}  by the  \( C_{\mathrm{q}} (\sqrt{s}) \)  
of the multiplicity distributions measured by the ATLAS and complemented with  the CMS 
measurements at  
\(\sqrt{s} =0.9,\ 2.36\), 
and \(7\)~TeV  \cite{CMS:2010qvf}
and   results of the  lower-energy experiments  NA22  
\cite{EHSNA22:1987syd},
UA1  
\cite{UA1:1989bou},  
and 
UA5 
\cite{UA5:1985hzd,UA5:1988gup}. 
The \( C_{\mathrm{q}} (\sqrt{s}) \)  calculations based on  ATLAS results  for  the
kinematic region  \( \mid\eta\mid < 2.5\), \(n_{\mathrm{ch}} \ge 2\) and  \(p_{\mathrm{T}} >100\)~MeV
are shown in  Fig.~\ref{fig_Cq_ATLAS_1}(a) and in Table~\ref{tab:Cq_ATLAS}.
The  ATLAS and CMS  results 
%are  
agree within the error range.
The values of  \( C_{\mathrm{q}} (\sqrt{s}) \)   with \(q=2, \dots, 5\) 
for all  experiments   
%increases linearly 
linearly increase 
with \( \log{\sqrt{s}}\) as illustrated by the fits in  
Fig.~\ref{fig_Cq_ATLAS_1}(a) and in Table~\ref{tab:Cq_fit_all} (phase space (A)).
Since, as mentioned above,  the KNO scaling requires  that \( C_{\mathrm{q}} (\sqrt{s}) \) 
be independent of energy,  we can state that the KNO scaling is violated at least for  the
full region of scaled multiplicity.

Figure~\ref{fig_Cq_ATLAS_1}(b)  shows for the first time the values of  \( C_{\mathrm{q}} (\sqrt{s}) \) 
calculated using multiplicity distributions  measured by  ATLAS for the kinematic region
\( \mid\eta\mid < 2.5\), \(n_{\mathrm{ch}} \ge 1\) and  \(p_{\mathrm{T}} >500\)~MeV.
Similarly,  
as in Fig.~\ref{fig_Cq_ATLAS_1}(a)  the values of  \( C_{\mathrm{q}} (\sqrt{s}) \) 
linearly increase with \( \log{\sqrt{s}}\)
% YK
as the function
\begin{equation}
%\nonumber
\label{eq_Cq_func}
 C_{\mathrm{q}} (\sqrt{s}) = \alpha +\beta \sqrt{s}.
\end{equation}
The  results of the fit are presented in Table~\ref{tab:Cq_fit_all} (phase space (B)).
The \( C_{\mathrm{q}} \) values at \(\sqrt{s} = 2.36\)~TeV  in Fig.~\ref{fig_Cq_ATLAS_1}(b) 
are much smaller  than those for other energies.
This  is because the  region of primary charged-particle multiplicity distributions at  \(2.36\)~TeV is  
smaller (up to \(z \approx 3.5\))  than that for higher  CM  energies (up to \(z \approx 9\)) 
\cite{Kulchitsky:2022gkm}.
Therefore,  the \( C_{\mathrm{q}} \) values at \(\sqrt{s} = 2.36\)~TeV  were not used in the fits. 
The  \( C_{\mathrm{q}} (\sqrt{s}) \)    for  \(p_{\mathrm{T}} >500\)~MeV  
has a  higher  
bias  (\( \alpha \))  and  slope (\( \beta \))   of the fits than those for   
the  minimum \(p_{\mathrm{T}}\) threshold, 
with 
the bias  increasing  from \(1.1\) at  \(q=2\) up to  \(2.1\) at  \(q=5\), 
and  
the slope   increasing from \(1.4\) at  \(q=2\) up to  \(2.6\) at  \(q=5\).
This is the result of  stronger interactions with  a higher \(p_{\mathrm{T}}\) threshold  in 
case (B) than in  case (A).
%%
% YK
This is  because the total transverse momentum, \(\sum p_{\mathrm{T}}\),
for events  with \(p_{\mathrm{T}} >500\)~MeV  (B) 
is higher than the same observable 
%\(\sum p_{\mathrm{T}}\) 
for events  with \(p_{\mathrm{T}} >100\)~MeV  (A).

Figure~\ref{fig_Cq_ATLAS_1}(c)   shows  moments  \(C_{\mathrm{q}}\)   for events with 
\(n_{\mathrm{ch}} \ge 2\), \(p_{\mathrm{T}} >100\)~MeV, 
 and for  \( z > 0.5 \)
without  the fraction of single and double diffraction events, which was accepted  by 
the ATLAS minimum-bias trigger
\cite{STDM-2010-01,STDM-2010-06,STDM-2014-19,STDM-2015-02,STDM-2015-17}.
In this case,  the values of  \( C_{\mathrm{q}} (\sqrt{s}) \)  are systematically higher than those for 
full distributions  with  \( z > 0\) and  show a similar linear increase with \( \log{\sqrt{s}}\)  
as 
%is 
 illustrated in   Fig.~\ref{fig_Cq_ATLAS_1}(c).
For multiplicity distributions for \( z > 1.0 \), 
 the  values of  \( C_{\mathrm{q}} (\sqrt{s}) \)  
at the highest energies,   
\( \sqrt{s} =7,\ 8\), and \(13\)~TeV, 
are  in agreement within error uncertainties,
as can be seen  in Fig.~\ref{fig_Cq_ATLAS_1}(c).
Therefore,  the energy independence of  the moments  of various orders can be considered as a confirmation
 of  the KNO scaling. 
This is in agreement with  the conclusion in  \cite{Kulchitsky:2022gkm} that the KNO scaling holds for 
the highest energies.

%______________________________________________________________________________
% Table 1
%______________________________________________________________________________
\begin{table*}[t!]
%\centering 
\caption{
The  normalised order-\(q\) moments,   \(C_{\mathrm{q}} (\sqrt{s}) \),    of the primary charged-particle
 multiplicity distributions  measured by the ATLAS Collaboration  for events  at center-of-mass energies 
\(\sqrt{s} = 0.9,\ 2.36,\ 7, 8\), and \(13\)~TeV 
for pseudorapidity region  \( \mid\eta\mid < 2.5\)  and  
for two different phase spaces \(n_{\mathrm{ch}} \ge 2\), \(p_{\mathrm{T}} >100\)~MeV
and \(n_{\mathrm{ch}} \ge 1\),  \(p_{\mathrm{T}} >500\)~MeV. 
}
\label{tab:Cq_ATLAS}
%\medskip
 \begin{tabular}{@{}r@{~}r@{~}c@{~}c@{~}c@{~}c@{}}
\hline
\hline

   \(\sqrt{s}\)
& 
\(p_{\mathrm{T}}^{\mathrm{min}}\) 
%\(p_{\mathrm{T}}>\) 
& \multicolumn{4}{c}{\(C_{\mathrm{q}}\)} 
\\
\cline{3-6}

   {\small [TeV]}
& {\small [MeV]}
& \(C_{\mathrm{2}}\) 
& \(C_{\mathrm{3}}\) 
& \(C_{\mathrm{4}}\) 
& \(C_{\mathrm{5}}\)  
\\
\hline

\multicolumn{6}{c}{} 
\\[-3mm]

 13	
 &\(100\)  
 & \(1.799\pm0.002^{+0.021}_{-0.016}\) 	%C2
 & \(4.491\pm0.008^{+0.119}_{-0.084}\) 	%C3
 & \(13.74\pm0.04^{+0.65}_{-0.52}\) 		%C4
 & \(48.49\pm0.15^{+1.52}_{-0.66}\) 		%C5
 \\[2mm]
 
 &\(500\) 
 & \(2.065\pm0.002^{+0.008}_{-0.007}\) 	%C2
 & \(6.209\pm0.009^{+0.058}_{-0.047}\) 	%C3
 & \(23.57\pm0.05^{+0.46}_{-0.34}\) 		%C4
 & \(105.77\pm0.29^{+3.05}_{-2.55}\) 	%C5 
 \\[2mm]
 
\hline
\multicolumn{6}{c}{} 
\\[-3mm]

8	
 &\(100\) 
 & \(1.741\pm0.001^{+0.030}_{-0.029}\) 	%C2
 & \(4.185\pm0.007^{+0.131}_{-0.126}\) 	%C3
 & \(12.32\pm0.03^{+0.55}_{-0.54}\) 		%C4
 & \(41.81\pm0.14^{+2.40}_{-2.40}\) 		%C5
 \\[2mm]
 
 &\(500\) 
 & \(2.000\pm0.001^{+0.030}_{-0.030}\) 	%C2
 & \(5.793\pm0.008^{+0.161}_{-0.157}\) 	%C3
 & \(21.12\pm0.04^{+0.83}_{-0.81}\) 		%C4
 & \(90.81\pm0.21^{+4.56}_{-4.44}\) 		%C5 
 \\[2mm]
 
\hline
\multicolumn{6}{c}{} 
\\[-3mm]

7	
 &\(100\) 
 & \(1.712\pm0.005^{+0.026}_{-0.016}\) 	%C2
 & \(4.025\pm0.022^{+0.124}_{-0.062}\) 	%C3
 & \(11.58\pm0.10^{+0.64}_{-0.37}\) 		%C4
 & \(38.43\pm0.43^{+3.24}_{-2.04}\) 		%C5
 
 \\[2mm]
 
 &\(500\) 
 & \(1.985\pm0.004^{+0.018}_{-0.015}\) 	%C2
 & \(5.701\pm0.020^{+0.127}_{-0.116}\) 	%C3
 & \(20.59\pm0.10^{+0.87}_{-0.88}\) 		%C4
 & \(87.64\pm0.52^{+6.90}_{-6.56}\) 		%C5  
 \\[2mm]
 
\hline
\multicolumn{6}{c}{} 
\\[-3mm]

2.36
 &\(500\)  
 & \(1.759\pm0.047^{+0.075}_{-0.070}\) 	%C2
 & \(4.082\pm0.227^{+0.352}_{-0.313}\) 	%C3
 & \(11.00\pm0.95^{+1.46}_{-1.25}\) 		%C4
 & \(32.43\pm3.84^{+5.94}_{-4.86}\) 		%C5 
 \\[2mm]
 
\hline
\multicolumn{6}{c}{} 
\\[-3mm]

0.9	
 &\(100\) 
 & \(1.530\pm0.015^{+0.028}_{-0.021}\) 	%C2
 & \(3.121\pm0.057^{+0.102}_{-0.076}\) 	%C3
 & \(7.77\pm0.21^{+0.39}_{-0.30}\) 		%C4
 & \(22.33\pm0.77^{+1.72}_{-1.36}\) 		%C5
 \\[2mm]
 
 &\(500\) 
 & \(1.752\pm0.011^{+0.016}_{-0.015}\) 	%C2
 & \(4.364\pm0.052^{+0.091}_{-0.093}\) 	%C3
 & \(13.72\pm0.24^{+0.53}_{-0.51}\) 		%C4
 & \(50.81\pm1.20^{+2.64}_{-2.61}\) 		%C5 
 \\[2mm]

\hline
\hline
\end{tabular}
\end{table*}

%______________________________________________________________________________

%______________________________________________________________________________
% Table 2
%______________________________________________________________________________
\begin{table*}[t!]
\centering 
\caption{ 
The fit parameters of energy dependence of the   distributions of  
% YK
normalised order-\(q\) moments 
\(C_{\mathrm{q}}(\sqrt{s}) \) 
% YK
in Eq.~(\ref{eq_Cq_func})
for two different  phase spaces: 
(A)  \( \mid\eta\mid < 2.5\),
\(n_{\mathrm{ch}} \ge 2\), \(p_{\mathrm{T}} >100\)~MeV, 
and
(B)  \( \mid\eta\mid < 2.5\),  \(n_{\mathrm{ch}} \ge 1\),   \(p_{\mathrm{T}} >500\)~MeV. 
}
\label{tab:Cq_fit_all}
%\medskip
 \begin{tabular}{cccc}
\hline
\hline

\(C_{\mathrm{q}}\) 
& 
Phase Space
& \( \alpha \) 
& \( \beta \)
%&   \(\chi^2 /{\mathrm{ndf}}\)
\\
\hline

\(C_{\mathrm{2}}\)  	& (A)  & \(1.54\pm0.01\)  & \(0.200\pm0.015\)   \\
%& \(55/16\) 	 \\
 						& (B)  & \(1.76\pm0.02\)  & \(0.271\pm0.017\)   \\[1mm]
% 						& \(0.2/2\) 	 \\[1mm]
 
\(C_{\mathrm{3}}\)  	& (A)  & \(3.31\pm0.03\)  & \(0.907\pm0.031\)   \\
%& \(10/9\) 	 \\
 						& (B)  & \(4.42\pm0.10\)  & \(1.59\pm0.10\)   \\[1mm]
 						%& \(0.5/2\) 	\\[1mm]
 
\(C_{\mathrm{4}}\)  	& (A)  & \(8.86\pm0.18\)  & \(3.40\pm0.13\)   \\
%& \(15/9\) 	 \\
 						& (B)  & \(14.0\pm0.5\)  & \(8.45\pm0.56\)   \\[1mm]
 						%& \(1.0/2\) 	\\[1mm]
 
\(C_{\mathrm{5}}\)  	& (A)  & \(25.2\pm0.7\)  & \(17.6\pm1.4\)  \\
% & \(7/8\) 	 \\
 						& (B)  & \(52.5\pm2.7\)  & \(46.5\pm3.2\)   \\[1mm]
 						%& \(1.4/2\) 	\\[1mm]
 
\hline
\hline
\end{tabular}
\end{table*}
%______________________________________________________________________________

The results of  studying \(C_{\mathrm{q}}(\sqrt{s}) \)   of primarily charged-particle multiplicity 
distributions using the results of the ATLAS  at the LHC are presented.
The  normalised  order-\(q\)  moments  are sensitive to the KNO scaling.
The \(C_{\mathrm{q}}(\sqrt{s}) \)  from  the ATLAS, CMS, and low-energy experiments are compared 
for the kinematic region with   an  absolute pseudorapidity less than \(2.5\)  of \( \sqrt{s} \) from  \(0.2\) to \(13\)~TeV.  
For  the full-scale multiplicity region,   the \( C_{\mathrm{q}} (\sqrt{s}) \)  moments show a linear 
increase with   \( \sqrt{s} \),  therefore indicating that KNO scaling  is  violated. 
The \( C_{\mathrm{q}}  (\sqrt{s}) \) of the  scaled multiplicity  larger  
than one average multiplicity    is  constant for the highest energies.
%
%______________________________________________________________________________
% YK
Thus, for the first time, it can be concluded that for  \( \mid\eta\mid < 2.5 \) and \( z >1\),
the KNO scaling is valid.
Previously, the KNO scaling validity  at LHC energies was observed for \(\mid\eta\mid < 0.5\)  
by CMS \cite{CMS:2010qvf}.
%______________________________________________________________________________

We thank  the ATLAS Collaboration for the excellent experimental results 
%which 
that
were used for this analysis.
Our special  thanks go to Edward~K.~Sarkisyan-Grinbaum  and   Stanislav~Tokar  
for several productive discussions.

\newpage
%\printbibliography
%\end{document}

%____________________________

%_________________________________________________________________
%
%\end{document}
%_________________________________________________________________
%_________________________________________________________________
%_________________________________________________________________

% $ biblatex auxiliary file $
% $ biblatex bbl format version 3.1 $
% Do not modify the above lines!
%
% This is an auxiliary file used by the 'biblatex' package.
% This file may safely be deleted. It will be recreated as
% required.
%

%\input{sn-article.bbl}
%\input{Kulchitsky_KNO_EPJC.bbl}

%_____________________________________________________________________

%\bibliography{Kulchitsky_KNO_EPJC.bib}

\begin{thebibliography}{23}
%____________________________

% 1
%\bibitem{Kulchitsky:2023tcq} 
%Y.~A.~Kulchitsky,  
%Probe of soft-QCD in minimum bias events of pp collisions with the ATLAS at the LHC,
%arXiv: 2307.03925 [hep-ex].

\begin{comment}
@article{Kulchitsky:2023tcq,
    author = "Kulchitsky, Yuri A.",
    title = "{Probe of soft-QCD in minimum bias events of pp collisions with the ATLAS at the LHC}",
    eprint = "2307.03925",
    archivePrefix = "arXiv",
    primaryClass = "hep-ex",
    month = "7",
    year = "2023"
}
\end{comment}

% 2
\bibitem{Polyakov:1970lyy} 
A.~M.~Polyakov, A Similarity hypothesis in the strong interactions. 
1.\ Multiple hadron production in \(e^{+} e^{-}\) annihilation, 
Zh.\ Eksp.\ Teor.\ Fiz.\ {\textbf 59} (1970) 542.

\begin{comment}
@article{Polyakov:1970lyy,
    author = "Polyakov, A. M.",
    title = "{A Similarity hypothesis in the strong interactions. 
    1. Multiple hadron production in \( e^+ e^- \) annihilation}",
    journal = "Zh. Eksp. Teor. Fiz.",
    volume = "59",
    pages = "542--552",
    year = "1970"
}
\end{comment}

%3
\bibitem{Koba:1972ng} 
 Z.~Koba, H.~B.~Nielsen and P.~Olesen, Scaling of multiplicity distributions in high-energy hadron collisions, 
 Nucl.\ Phys.\ B {\textbf 40} (1972) 317.
 
\begin{comment}
@article{Koba:1972ng,
    author = "Koba, Z. and Nielsen, Holger Bech and Olesen, P.",
    title = "{Scaling of multiplicity distributions in high-energy hadron collisions}",
    doi = "10.1016/0550-3213(72)90551-2",
    journal = "Nucl. Phys. B",
    volume = "40",
    pages = "317--334",
    year = "1972"
}
\end{comment}

% 4
\bibitem{Proceedings:1973mza}
Z.~Koba, Multi-body phenomena in strong interactions – description of hadronic multi-body final states, 
%(p. 171) 
in Proceedings  CERN-JINR School of Physics, Ebeltoft, Denmark, 17-13 June 1973.
%, CERN Yellow Reports: School Proceedings (1973).

\begin{comment}
@article{Proceedings:1973mza,
    author = "Koba, Z.",
    title = "{Multi-body phenomena in strong interactions -- description of hadronic multi-body final states,  
    (p. 171) in Proceedings CERN-JINR School of Physics, Ebeltoft, Denmark, 17-13 June 1973}",
    reportNumber = "CERN-73-12",
    doi = "10.5170/CERN-1973-012",
    series = "CERN Yellow Reports: School Proceedings",
    month = "9",
    year = "1973"
}
\end{comment}

% 5
\bibitem{Feynman:1969ej}
R.~P.~Feynman, Very high-energy collisions of hadrons, 
Phys.\ Rev.\ Lett.\ {\textbf 23} (1969) 1415, ed.\ by L.~M.~Brown.

\begin{comment}
@article{Feynman:1969ej,
    author = "Feynman, Richard P.",
    editor = "Brown, L. M.",
    title = "{Very high-energy collisions of hadrons}",
    reportNumber = "PRINT-69-2817",
    doi = "10.1103/PhysRevLett.23.1415",
    journal = "Phys. Rev. Lett.",
    volume = "23",
    pages = "1415--1417",
    year = "1969"
}
\end{comment}

% 6
\bibitem{Grosse-Oetringhaus:2009eis}
J.~F.~Grosse-Oetringhaus and K.~Reygers, Charged-Particle Multiplicity in Proton-Proton Collisions, 
J.\ Phys.\ G {\textbf 37} (2010) 083001, arXiv: 0912.0023 [hep-ex].

\begin{comment}
@article{Grosse-Oetringhaus:2009eis,
    author = "Grosse-Oetringhaus, Jan Fiete and Reygers, Klaus",
    title = "{Charged-Particle Multiplicity in Proton-Proton Collisions}",
    eprint = "0912.0023",
    archivePrefix = "arXiv",
    primaryClass = "hep-ex",
    doi = "10.1088/0954-3899/37/8/083001",
    journal = "J. Phys. G",
    volume = "37",
    pages = "083001",
    year = "2010"
}
\end{comment}

% 7
\bibitem{Kittel:2005fu}
W.~Kittel and E.~A.~De Wolf, Soft multihadron dynamics, 2005, isbn: 978-981-256-295-1.

\begin{comment}
@book{Kittel:2005fu,
    author = "Kittel, W. and De Wolf, E. A.",
    title = "{Soft multihadron dynamics}",
    isbn = "978-981-256-295-1",
    year = "2005"
}
\end{comment}


% 8
\bibitem{STDM-2010-01} 
ATLAS Collaboration, 
Charged-particle multiplicities in \(pp\) interactions at \(\sqrt{s} = 900\)~GeV measured with the ATLAS detector at the LHC, 
Phys.\ Lett.\ B {\textbf 688}  (2010) 21,
arXiv: 1003.3124 [hep-ex].

\begin{comment}
@Article{STDM-2010-01,
    author         = "{ATLAS Collaboration}",
    title          = "{Charged-particle multiplicities in \(pp\) interactions at \(\sqrt{s} = 900\,\textrm{GeV}\) 
    measured with the ATLAS detector at the LHC}",
    journal        = "Phys. Lett. B",
    volume         = "688",
    year           = "2010",
    pages          = "21",
    doi            = "10.1016/j.physletb.2010.03.064",
    reportNumber   = "CERN-PH-EP-2010-004",
    eprint         = "1003.3124",
    archivePrefix  = "arXiv",
    primaryClass   = "hep-ex",
}
\end{comment}

% 9
\bibitem{STDM-2010-06} 
ATLAS Collaboration,
Charged-particle multiplicities in \(pp\)  interactions measured with the ATLAS detector at the LHC,
New J.\ Phys.\ {\textbf 13} (2011) 053033, 
arXiv: 1012.5104 [hep-ex].

\begin{comment}
@Article{STDM-2010-06,
    author         = "{ATLAS Collaboration}",
    title          = "{Charged-particle multiplicities in \(pp\) interactions measured with 
    the ATLAS detector at the LHC}",
    journal        = "New J. Phys.",
    volume         = "13",
    year           = "2011",
    pages          = "053033",
    doi            = "10.1088/1367-2630/13/5/053033",
    reportNumber   = "CERN-PH-EP-2010-079",
    eprint         = "1012.5104",
    archivePrefix  = "arXiv",
    primaryClass   = "hep-ex",
}
\end{comment}

% 10
\bibitem{STDM-2014-19} 
ATLAS Collaboration,
Charged-particle distributions in \(pp\) interactions at \(\sqrt{s} = 8\)~TeV measured with the ATLAS detector,
Eur.\ Phys.\ J.\ C {\textbf 76} (2016) 403, 
arXiv: 1603.02439 [hep-ex].

\begin{comment}
@Article{STDM-2014-19,
    author         = "{ATLAS Collaboration}",
    title          = "{Charged-particle distributions in \(pp\) interactions at \(\sqrt{s} = 8\,\textrm{TeV}\) measured with the ATLAS detector}",
    journal        = "Eur. Phys. J. C",
    volume         = "76",
    year           = "2016",
    pages          = "403",
    doi            = "10.1140/epjc/s10052-016-4203-9",
    reportNumber   = "CERN-EP-2016-020",
    eprint         = "1603.02439",
    archivePrefix  = "arXiv",
    primaryClass   = "hep-ex",
}
\end{comment}

% 11
\bibitem{STDM-2015-02} 
ATLAS Collaboration, 
Charged-particle distributions in \(\sqrt{s} = 13\)~TeV \(pp\) interactions measured with the ATLAS detector at the LHC, 
Phys. Lett. B {\textbf 758} (2016) 67, 
arXiv: 1602.01633 [hep-ex].

\begin{comment}
@Article{STDM-2015-02,
    author         = "{ATLAS Collaboration}",
    title          = "{Charged-particle distributions in \(\sqrt{s} = 13\,\textrm{TeV}\) \(pp\) interactions 
    measured with the ATLAS detector at the LHC}",
    journal        = "Phys. Lett. B",
    volume         = "758",
    year           = "2016",
    pages          = "67",
    doi            = "10.1016/j.physletb.2016.04.050",
    reportNumber   = "CERN-EP-2016-014",
    eprint         = "1602.01633",
    archivePrefix  = "arXiv",
    primaryClass   = "hep-ex",
}
\end{comment}

% 12
\bibitem{STDM-2015-17} 
ATLAS Collaboration, 
Charged-particle distributions at low transverse momentum in \(\sqrt{s} = 13\)~TeV
\(pp\) interactions measured with the ATLAS detector at the LHC, 
Eur.\ Phys.\ J.\ C {\textbf 76} (2016) 502,
arXiv: 1606.01133 [hep-ex].

\begin{comment}
@Article{STDM-2015-17,
    author         = "{ATLAS Collaboration}",
    title          = "{Charged-particle distributions at low transverse momentum in 
    \(\sqrt{s} = 13\,\textrm{TeV}\) \(pp\) interactions measured with the ATLAS detector at the LHC}",
    journal        = "Eur. Phys. J. C",
    volume         = "76",
    year           = "2016",
    pages          = "502",
    doi            = "10.1140/epjc/s10052-016-4335-y",
    reportNumber   = "CERN-EP-2016-099",
    eprint         = "1606.01133",
    archivePrefix  = "arXiv",
    primaryClass   = "hep-ex",
}
\end{comment}

% 13
\bibitem{PERF-2007-01} 
ATLAS Collaboration, 
The ATLAS Experiment at the CERN Large Hadron Collider,
JINST {\textbf 3} (2008) S08003.

\begin{comment}
@Article{PERF-2007-01,
    author         = "{ATLAS Collaboration}",
    title          = "{The ATLAS Experiment at the CERN Large Hadron Collider}",
    journal        = "JINST",
    volume         = "3",
    year           = "2008",
    pages          = "S08003",
    doi            = "10.1088/1748-0221/3/08/S08003",
    primaryClass   = "hep-ex",
}
\end{comment}

% 14
\bibitem{Evans:2008zzb} 
L.~Evans and P.~Bryant, 
LHC Machine, 
JINST {\textbf 3} S08001 (2008).

\begin{comment}
@Article{Evans:2008zzb,
      author         = "Evans, Lyndon and Bryant, Philip",
      title          = "{LHC Machine}",
      journal        = "JINST",
      volume         = "3",
      pages          = "S08001",
      doi            = "10.1088/1748-0221/3/08/S08001",
      year           = "2008",
      SLACcitation   = "%%CITATION = JINST,3,S08001;%%",
}
\end{comment}

% 15
\bibitem{Kulchitsky:2022gkm} 
Y.~Kulchitsky and P.~Tsiareshka, Study of the KNO scaling in \(pp\) collisions at \(\sqrt{s}\)  
from 0.9 to 13~TeV using results of the ATLAS at the LHC, 
Eur.\ Phys.\ J.\ C {\textbf 82} (2022) 462, arXiv: 2202.06697 [hep-ex].

\begin{comment}
@article{Kulchitsky:2022gkm,
    author = "Kulchitsky, Yuri and Tsiareshka, Pavel",
    title = "{Study of the KNO scaling in $ pp$ collisions at $\sqrt{s}$ 
    from 0.9 to 13~TeV using results of the ATLAS at the LHC}",
    eprint = "2202.06697",
    archivePrefix = "arXiv",
    primaryClass = "hep-ex",
    doi = "10.1140/epjc/s10052-022-10420-y",
    journal = "Eur. Phys. J. C",
    volume = "82",
    number = "5",
    pages = "462",
    year = "2022"
}
\end{comment}


% 16
\bibitem{CMS:2010qvf} 
CMS Collaboration, 
Charged Particle Multiplicities in \(pp\) Interactions at \( \sqrt{s}=0.9\), 2.36 and 7~TeV,
JHEP {\textbf 01} (2011) 079, 
arXiv: 1011.5531 [hep-ex].

\begin{comment}
@article{CMS:2010qvf,
    author         = "{CMS Collaboration}",
    title = "{Charged Particle Multiplicities in $pp$ Interactions at $\sqrt{s}=0.9$, 2.36, and 7~TeV}",
    eprint = "1011.5531",
    archivePrefix = "arXiv",
    primaryClass = "hep-ex",
    reportNumber = "CERN-PH-EP-2010-048, CMS-QCD-10-004",
    doi = "10.1007/JHEP01(2011)079",
    journal = "JHEP",
    volume = "01",
    pages = "079",
    year = "2011"
}
\end{comment}

% 17
\bibitem{ALICE:2010cin} 
ALICE Collaboration, 
Charged-particle multiplicity measurement in proton-proton collisions at
\(\sqrt{s} = 0.9\) and 2.36~TeV with ALICE at LHC, 
Eur.\ Phys.\ J.\ C {\textbf 68} (2010) 89,
arXiv: 1004.3034 [hep-ex].

\begin{comment}
@article{ALICE:2010cin,
    author         = "{ALICE Collaboration}",
    title = "{Charged-particle multiplicity measurement in proton-proton collisions at 
    $\sqrt{s} = 0.9$ and 2.36~TeV with ALICE at LHC}",
    eprint = "1004.3034",
    archivePrefix = "arXiv",
    primaryClass = "hep-ex",
    doi = "10.1140/epjc/s10052-010-1339-x",
    journal = "Eur. Phys. J. C",
    volume = "68",
    pages = "89--108",
    year = "2010"
}
\end{comment}

% 18
\bibitem{ALICE:2015olq} 
ALICE Collaboration, 
Charged-particle multiplicities in proton\textendash{}proton collisions at \(\sqrt{s} = 0.9\) to 8~TeV,
Eur.\ Phys.\ J.\ C {\textbf 77} (2017) 33, 
arXiv: 1509.07541 [nucl-ex].

\begin{comment}
@article{ALICE:2015olq,
    author         = "{ALICE Collaboration}",
    collaboration = "ALICE",
    title = "{Charged-particle multiplicities in proton\textendash{}proton collisions at 
    $\sqrt{s} = 0.9$ to 8~TeV}",
    eprint = "1509.07541",
    archivePrefix = "arXiv",
    primaryClass = "nucl-ex",
    reportNumber = "CERN-PH-EP-2015-259",
    doi = "10.1140/epjc/s10052-016-4571-1",
    journal = "Eur. Phys. J. C",
    volume = "77",
    number = "1",
    pages = "33",
    year = "2017"
}
\end{comment}

% 19
\bibitem{ALICE:2022xip} 
ALICE Collaboration, 
Multiplicity dependence of charged-particle production in  \(pp\), \(p\)-\(Pb\), \(Xe\)-\(Xe\) and \(Pb\)-\(Pb\) 
collisions at the LHC,
CERN-EP-2022-266, Geneva, 2022, 
arXiv: 2211.15326 [nucl-ex].

\begin{comment}
@article{ALICE:2022xip,
    author         = "{{ALICE~Collaboration}}",
    collaboration = "ALICE",
    title = "{Multiplicity dependence of charged-particle production in 
    \(pp\), \(p\)-\(Pb\), \(Xe\)-\(Xe\) and \(Pb\)-\(Pb\) collisions at the LHC}",
    eprint = "2211.15326",
    archivePrefix = "arXiv",
    primaryClass = "nucl-ex",
    reportNumber = "CERN-EP-2022-266",
    month = "11",
    year = "2022"
}
\end{comment}

% 20
\bibitem{Fan:2022bbp} 
ALICE Collaboration, 
F.~Fan for the Collaboration, 
Particle production as a function of underlying-event activity and very forward energy with ALICE,
Eur.\ Phys.\ J.\ Web Conf.\  {\textbf 276 } (2023) 01009, 
arXiv: 2208.11348 [nucl-ex].

\begin{comment}
@article{Fan:2022bbp,
    author = "Fan, Feng",
    collaboration = "ALICE",
    title = "{Particle production as a function of underlying-event activity and very forward energy with ALICE}",
    eprint = "2208.11348",
    archivePrefix = "arXiv",
    primaryClass = "nucl-ex",
    doi = "10.1051/epjconf/202327601009",
    journal = "EPJ Web Conf.",
    volume = "276",
    pages = "01009",
    year = "2023"
}
\end{comment}

% 21 
\bibitem{EHSNA22:1987syd} 
EHS/NA22 Collaboration, Phase Space Dependence of the Multiplicity Distribution in \( \pi^+ p\) and \(pp\) Collisions at 250-GeV/c,
Z.\ Phys.\ C {\textbf 37} (1988) 215.

\begin{comment}
@article{EHSNA22:1987syd,
    author         = "{{EHS/NA22~Collaboration}}",
    collaboration = "EHS/NA22",
    title = "{Phase Space Dependence of the Multiplicity Distribution in 
    $\pi^+ p$ and $p p$ Collisions at 250-{GeV}/$c$}",
    reportNumber = "HEN-284",
    doi = "10.1007/BF01579907",
    journal = "Z. Phys. C",
    volume = "37",
    pages = "215",
    year = "1988"
}
\end{comment}

% 22
\bibitem{UA1:1989bou} 
UA1 Collaboration, 
A Study of the General Characteristics of \(p\bar{p}\) Collisions at \(\sqrt{s} = 0.2\)-TeV to \(0.9\)-TeV,
Nucl.\ Phys.\ B {\textbf 335} (1990) 261.

\begin{comment}
@article{UA1:1989bou,
    author         = "{UA1 Collaboration}",
    collaboration = "UA1",
    title = "{A Study of the General Characteristics of $p\bar{p}$ Collisions at 
    $\sqrt{s} = 0.2$-TeV to 0.9-TeV}",
    reportNumber = "CERN-EP-89-85",
    doi = "10.1016/0550-3213(90)90493-W",
    journal = "Nucl. Phys. B",
    volume = "335",
    pages = "261--287",
    year = "1990"
}
\end{comment}

% 23
\bibitem{UA5:1985hzd} 
UA5 Collaboration, 
An Investigation of Multiplicity Distributions in Different Pseudorapidity Intervals in anti-p p Reactions at a CMS Energy of 540-GeV, 
Phys.\ Lett.\ B {\textbf 160} (1985) 193.

\begin{comment}
@article{UA5:1985hzd,
    author         = "{UA5 Collaboration}",
    collaboration = "UA5",
    title = "{An Investigation of Multiplicity Distributions in Different Pseudorapidity Intervals in anti-p p 
    Reactions at a CMS Energy of 540-GeV}",
    reportNumber = "CERN-EP/85-61",
    doi = "10.1016/0370-2693(85)91491-1",
    journal = "Phys. Lett. B",
    volume = "160",
    pages = "193--198",
    year = "1985"
}
\end{comment}

% 24
\bibitem{UA5:1988gup} 
UA5 Collaboration, 
Charged Particle Multiplicity Distributions at 200-GeV and 900-GeV Center-Of-Mass Energy,
Z.\ Phys.\ C {\textbf 43} (1989) 357, 
ed.\ by R.\ Kotthaus and J.\ H.\ Kuhn.

\begin{comment}
@article{UA5:1988gup,
    author         = "{UA5 Collaboration}",
    editor = "Kotthaus, R. and Kuhn, Johann H.",
    collaboration = "UA5",
    title = "{Charged Particle Multiplicity Distributions at 200-GeV and 900-GeV 
    Center-Of-Mass Energy}",
    reportNumber = "CERN-EP-88-172",
    doi = "10.1007/BF01506531",
    journal = "Z. Phys. C",
    volume = "43",
    pages = "357",
    year = "1989"
}
\end{comment}


%_________________________________________________________________
\end{thebibliography}

%\input sn-sample-bib.tex%

\end{document}